\documentclass[11pt,english]{article}
\usepackage[T1]{fontenc}
\usepackage{color}
\usepackage{amsmath}
\usepackage{amssymb}
\usepackage{float}
\usepackage{setspace}
\usepackage{esint}
\usepackage{multirow}
\makeatletter
\usepackage{physics}
\usepackage{amsthm}
\usepackage{graphicx}
\usepackage{amsfonts}
\usepackage{slashed}
\usepackage{afterpage}
\usepackage{cite}
\usepackage{bm}

\definecolor{darkgreen}{cmyk}{1,0,1,0}

\usepackage{stackengine}

\stackMath

\usepackage[margin=1in]{geometry}

\usepackage{babel}

\makeatother

\usepackage{babel}
\begin{document}

	\title{The geometry of physical observables}
	\author{S. Farnsworth}
	\maketitle
	 \begin{center}
   Max Planck Institute for Gravitational Physics (Albert Einstein Institute), Germany.
	\end{center} 
	\begin{abstract}
		
		Jordan algebras were first introduced in an effort to restructure quantum mechanics purely in terms of physical observables. In this paper we explain why, if one attempts to reformulate the internal structure of the standard model of particle physics geometrically, one  arrives naturally at a discrete internal geometry that is coordinatized by a Jordan algebra.


	\end{abstract}
		\tableofcontents

	\section{Introduction}
	\label{Sec_Introduction}

Ever since it was first discovered that gravity corresponds to the geometry of four dimensional spacetime, physicists have	wondered if all known fundamental fields, including gravity, might be unified by interpreting them as describing the geometry of some appropriately extended spacetime.  Perhaps the most famous implementation of this strategy is that of `Kaluza-Klein theory', which imagines spacetime as a product space $M=M_4\times M_{int}$, where $M_4$ is our familiar  four dimensional spacetime, while $M_{int}$ is another `internal' space that is small and compact, but which otherwise is  an ordinary smooth Riemannian manifold. 	The idea is that the particular compactified structure of the  internal dimensions  determines the particle content and gauge symmetries that we detect in low energy particle 	experiments. This idea is beautiful, but  it is unfortunately difficult to implement and in particular to stabilize the extra internal dimensions. At its heart, the problem with this approach is tightly linked to the assumption that the internal space should be a smooth manifold, or correspondingly that it has many potentially unstable continuous deformations. In this paper we drop this assumption, and explore the possibility of replacing the smooth internal manifolds  of Kaluza-Klein theory with discrete internal spaces that are non-dynamical (i.e. not Riemannian). 
	

If a discrete internal geometry underlies the structure of the standard model of particle physics, then what should it ``look'' like? A hint can be found in the usual description of  Weyl fermions within the familiar framework of quantum field theory. Unlike in the traditional approach to quantum mechanics, the position of  a particle in quantum field theory is treated as a continuous label rather than as an operator. That is, we can think of the field $\Psi(x)$ as a spinor valued object living over the point labelled  `$x$' in spacetime. In reality, though, we don't just have a single species of fermion. Instead the standard model describes a collection of $48$ Weyl fermions $\Psi_k(x)$ indexed by the discrete label $k = 1,...,48$. The usual approach is to think of these as $48$ independent spinor fields, each of which lives over four dimensional spacetime. Notice, however,
	that `$x$' and `$k$' are both just labels, and so it is natural to wonder why it is that we only  treat `$x$' geometrically. An alternative approach would be to treat both labels on exactly the same geometric footing. From such a perspective $\Psi_k(x)$ would be viewed as a single spinor field living over an extended `spacetime', which consists of two parts: an ordinary, smooth four dimensional manifold (corresponding to the label $x$) and an internal
	geometric space (corresponding to the discrete label $k = 1,...,48$).

	In order to see more precisely how a discrete species index `$k$' 	  might correspond to a discrete `internal' geometry, it is useful to think about how geometries are usually coordinatized. In the familiar setting of Riemannian geometry, the points of a  manifold  are featureless with no internal structure whatsoever.	In this case, a manifold `$M$' is  coordinatized  by functions `$f$', each of which smoothly  associates to each point $x\in M$ a single  number `$f(x)$' (i.e. these are the kinds of functions that are usually  denoted  by `$x^\mu(x)$'). The set of coordinate functions defined on a manifold form a coordinate algebra pointwise\footnote{This means that the addition and multiplication of functions is defined locally at each point on the manifold, i.e. $(f+g)(x) = f(x)+g(x)$  and $fg(x) = f(x)g(x)$ for $f,g\in A$.}, which we denote $A= C^\infty(M,\mathbb{C})$. For a spin manifold (i.e. a manifold that admits a spinor bundle), one way of viewing coordinate functions, is as  `position-type' operators that act  pointwise on spinor fields. That is, given a coordinate function $f\in A$, and an element  $\Psi$ in the Hilbert space of square integrable Dirac spinors $H=L^2(M,S)$, one defines the following pointwise action: $(f\Psi)(x) = f(x)\Psi(x)$.  The eigenvectors of the coordinate algebra are then the wavefunctions $\Psi\in H$ which are perfectly localized at any particular point `$x$' on the manifold. In other words, the underlying manifold is recovered as a spectrum of the coordinate algebra of  `position type' operators. The intuitive picture is one in which particles themselves are  in some sense  the instantiation  of the points (i.e. representations) on the manifold (i.e. the coordinate algebra).
		
	How does the above picture extend to the  case in which 
	the spinor fields $\Psi\in H$ are equipped with an additional `discrete' index `$k$'? To give this new index   the same geometric meaning as   the continuous coordinate label `$x$',  the coordinate algebra of `position type' observables will need to be extended to ``see'' the internal space. That is, if we wish to imbue the points on a manifold with  	some additional internal structure, then the usual coordinatization will no longer suffice because the 	algebra of coordinate functions must now encode information not only about the location of points, but also the internal `state' of each
 	point. 	A natural conjecture is that the geometry should take the form of a  product space,  coordinatized by an algebra of the form $A = C^\infty(M,\mathbb{C})\otimes A_F$, where $A_F$ is a finite dimensional (and possibly discrete) algebra that encodes the details of the internal space\footnote{
The coordinate algebra $A = C^\infty(M,\mathbb{C})\otimes A_F = C^\infty(M,A_F)$  smoothly associates  to each point on the manifold `$M$' an element of a finite dimensional algebra $A_F$, rather than an element of the algebra of complex numbers $\mathbb{C}$. The idea is that $A_F$ holds internal information about the manifold at each point.}.  	 Similarly, the spinor fields $\Psi_k(x)$ form a product Hilbert space denoted by $H = L^2(M,S)\otimes H_F$, where $H_F$ is a complex vector space on which $A_F$ is faithfully represented (and to which the index `k' is associated). In this setting, the `points' of the geometry can once again be associated with the spectrum or `states' of the coordinate algebra (or, following the GNS theorem, as the irreducible representations of the coorindate algebra). 
 
A remarkable feature of the above picture,  is that it  automatically presents a unified description of the four fundamental forces. In particular, general coordinate transformations of the coordinate algebra $A = C^\infty(M,A_F)$ correspond, not only to diffeomorphisms along the manifold $M$, but also to local rotations in the finite dimensional algebra $A_F$ at each point (i.e. local `gauge' transformations). The key dream of Kaluza-Klein theory appears to be   recovered, while the need for   any compactification scheme is avoided. The price that is paid, is that  the construction necessarily sits outside the usual framework of Riemannian geometry (the coordinate algebra is something more general). In order to make complete sense of the discrete internal geometry that underlies the standard model of particle physics, a number of geometric notions beyond that of  coordinate algebras and Hilbert spaces of spinors are  required, including an appropriate analogue of vector fields, differential forms, Dirac operators, Hochschild cohomology, Clifford representations of forms, and much more. Futhermore, in order to completely capture the structure  of the standard model of particle physics, the details of the Higgs sector must be filled in, and appropriate dynamics have to be determined.

An attempt at fully constructing the discrete internal geometry corresponding to the standard model of particle physics	has already been made within the framework of noncommutative differential geometry~\cite{Connes1994}.	In this approach the underlying internal topology is assumed to be coordinatized by a noncommutative matrix algebra, while metric information is encoded by a  generalized  Dirac operator. Remarkably, a   noncommutative geometry has been found, which captures (almost) all of the particle content and symmetries of the standard model, while providing new geometric meaning to a number of otherwise unexplained patterns and features observed in experiment~\cite{Chamseddine_2007}. 
As a key example, the standard model Higgs and gauge fields are unified within the construction, with the Higgs gaining new meaning as a connection on the discrete internal space. Furthermore, there are a number of geometric properties that generalize naturally from Riemannian geometry, and which in the noncommutative setting  place strict and phenomenologically accurate constraints on the particle content of the model.  
As an example, because the fermions in the model arise as irreducible representations of the coordinate algebra, the  matter content ends up being constrained by the representation theory for finite, noncommutative algebras. Remarkably, this alone
restricts  to those representations that are actually observed in experiment (singlet and fundamental).
	 
Despite the many intriguing features of the noncommutative geometric approach to particle theory,  the  noncommutative geometry that most closely captures the particle content of the standard model  does suffer from a number of problems and poorly understood details. These can best be seen where field content is added to or  removed from the construction  by hand  in order to match phenomenology. The so called `massless-photon'\footnote{In more recent work the mass-less photon condition has been replaced with the `second order condition', which does have geometric meaning~\cite{Boyle_2014}.} and `unimodularity' conditions~\cite{Chamseddine_2007}, for example, are employed to remove unwanted scalar and gauge bosons respectively, while  a scalar singlet has been introduced by hand in order to stabilize the Higgs mass at 125GeV~\cite{Chamseddine_2012}. Similarly, the three fermion generations remain as an unexplained input~\cite{Chamseddine_2008}.
 For the most part the ad-hoc addition and removal of particle content, while not having clear geometric justification, is not  in general inconsistent with the rules of the underlying geometry. As explained in~\cite{boyle2019standard}, however, this is unfortunately not always the case. More seriously, the construction suffers from what has been termed a `fermion quadrupling' problem, in which there are four times as many fermions in the model as observed experimentally~\cite{Lizzi_1997}. These unwanted states are able to be projected out~\cite{Barrett_2007}, but doing so results in a space of spinors that is no longer compatible with the required noncommutative coordinate algebra, in the sense that the set of standard model fermions is not closed under the action of the noncommutative coordinate functions~\cite{boyle2019standard}.

The inconsistencies  suffered by the noncommutative standard model do not necessarily indicate that there is a problem with the discrete internal geometric approach in general, but rather that there are problems with the specific geometry that has been proposed to match the details of the standard model. This is the position   taken in the current paper(as well as elsewhere in the literature~\cite{Chamseddine_2013,Devastato_2014}). When generalizing from Riemannian geometry to describe finite and discrete internal spaces there are often multiple `good' choices with regards to which properties should be kept or extended. These choices are important when modelling physical systems, and can lead to over- and under-constraint, both of which are seen in the noncommutative standard model. A particularly interesting   example of artificial over-constraint that is perhaps not so  obvious, is the axiomatic restriction to associative coordinate algebras~\cite{Chamseddine_2008}. Not only does this restriction automatically  exclude the possibility of exploring the geometry of gauge theories with exceptional symmetry~\cite{farnsworth2013nonassociative,Dubois_Violette_2016}, but it may ultimately also exclude those  geometries that are of most direct interest to physics  (including the standard model). In particular, while not compatible with associative, noncommutaive coordinate algebras, it turns out that the standard model fermions are compatible with the representation of a certain class of nonassociative algebras known as real Jordan algebras~\cite{boyle2019standard}. It is natural to wonder therefore if the internal structure underlying the standard model of particle physics might more accurately  be captured by a real `Jordan' geometry rather than a noncommutative geometry.

In this paper we provide  conceptual justification for considering internal geometries  coordinatized by real Jordan algebras if one is specifically interested in constructing gauge theories (coupled to Einstein-Hilbert gravity). Of particular concern is that the
reconstruction of Riemannian geometries appears to depend on $C^*$-algebras of complex coordinate functions $A=C^\infty(M,\mathbb{C})$ (viewed abstractly). It might therefore  seem at first glance that restricting attention to real Jordan coordinate algebras would explicitly exclude Riemannian geometry, let alone any interesting extensions that are capable to accurately capturing the internal details of the standard model.  Fortunately this appears not  to be the case. We explain how the  differential topological information encoded by a complex, noncommutative coordinate algebra is actually contained within its maximal real Jordan sub-algebra of selfadjoint elements. Furthermore, we explain how the physically relevant geometric quantities (i.e. symmetries, fermion representations, smooth structure, and distance measurements), continue to make sense when restricting attention to real Jordan coordinate algebras. We also provide an explicit construction of the differential calculi that will be required for eventually constructing the dynamics of geometries coordinatized by Jordan algebras.

This paper is organized as follows: in section \ref{Section_NCGSM} we briefly explore the noncommutative topology that most closely captures the internal fermion and gauge structure of the standard model of particle physics. We explain why such a construction  is not able to capture the full content of the physics we observe in experiment and explain why topological spaces coordinatized by  `Jordan' algebras naturally  circumvent this  same impediment. In section~\ref{jordsec} we explain how the physically interesting data contained in an associative coordinate algebra is fully contained within its maximal Jordan subalgebra. In particular, in section~\ref{Section_algebras_topology} we explain how the full topological data of a complex  $C^*$-algebra $A$ is encoded entirely within the set of selfadjoint elements of $A$, which form a real Jordan algebra $A_{sa}$. In section \ref{secjorddist} we explain how distance measurements on a geometry coordinatized by a complex $C^*$-algebra $A$ are able to be computed while only making reference to the maximal Jordan sub algebra $A_{sa}$. In section~\ref{sectjordsym} we explain how the symmetries of a geometry coordinatized by a complex $C^*$-algebra arise as (roughly) the automorphisms of the maximal real Jordan subalgebra $A_{sa}$ of $A$. In section~\ref{diffstructure} we explain how the differential structure encoded by an associative $C^*$-algebra is contained entirely within its maximal real Jordan subalgebra. We also explicitly construct the differential calculus over $A_{sa}$ that will be required for describing the dynamics of realistic gauge theories. In section \ref{seccon} we conclude, and explain what the next immediate steps are that will be required in order to explicitly construct the real Jordan geometry that most closely captures the full internal structure of the standard model of particle physics and explore its dynamics.

\section{The topology of the standard model}

\label{Section_NCGSM}

In theories with small extra dimensions, the usual approach is to imagine that spacetime is a product geometry, consisting of two parts $M = M_c\times M_F$. The first part $M_c$ is the familiar four dimensional, smooth, pseudo-Riemannian manifold that we observe and live in, while the second part $M_F$ is a manifold that is smooth and compact, but small enough  that  we don't yet observe it directly.  If the metric $g$ on the total space $M$ is assumed not to depend at all on the  internal part of the manifold $M_F$, then its components can be thought of as four dimensional fields living over the large `external' space $M_c$. The particular details of the geometry then encode information about gravity, as well as all of the fundamental fields and symmetries of particle theory.  In this paper we follow a similar approach,  with the main distinction being that we replace the internal space $M_F$  by a geometry that is instead finite and discrete. The usual hypothosis taken in the literature is that if such a discrete internal space is to exist, then it should be coordinatized by a noncommutative $C^*$-algebra.  Unfortunately the noncommutative geometry that most closely captures the particle content and symmetries of the standard model runs into a key  problem that appears difficult to resolve within the associative, noncommutative setting~\cite{boyle2019standard}. As we review in this section, this problem is neatly avoided by geometries coordinatized by Jordan algebras, which is a key motivation for this work. 

The `noncommutative' geometry that most closely captures the underlying structure of the  standard model of particle physics is constructed as a product geometry, coordinatized by an algebra of the form ${\cal A}={\cal A}_{c}\otimes{\cal A}_{F}$, bi-represented on a vector space of the form ${\cal H}={\cal H}_{c}\otimes{\cal H}_{F}$. Here the pair $\{{\cal A}_{c},{\cal H}_{c}\}$, capture the differential topological data of the external $4D$ spin-manifold `$M$' in which we live. In this case ${\cal A}_{c} = C^\infty(M,\mathbb{R})$ is the algebra of smooth functions on $M$, while ${\cal H}_{c} = L^2(M,S)$ is the Hilbert space of square integrable Dirac spinors defined on $M$. For the finite space the topology is captured by the finite dimensional algebra $A_F=\mathbb{C}\oplus\mathbb{H}\oplus M_3(\mathbb{C})$, where $\mathbb{C}$ is the algebra of complex numbers, $\mathbb{H}$ is the algebra of quaternions, and $M_3(\mathbb{C})$ is the algebra of $3\times 3$ complex matrices. This internal coordinate  algebra is chosen because its automorphism group corresponds (roughly) to the local gauge group of the standard model (deeper geometric motivation has subsequently been sought~\cite{Chamseddine_2008}). The algebra $A_F$ is represented on the $96$-dimensional complex Hilbert space $H_F = \mathbb{C}^{96}$, where $96$ corresponds to the total number of fermionic degrees of  freedom in the standard model of particle physics (after including particles and anti-particles,
left and right chiralities, right-handed neutrinos, 3 colors, and 3 families).


The representation of the finite dimensional algebra $A_F$ on $H_F$ is chosen to reproduce the particular details of the standard model. We explain this representation now, focusing for brevity on a single generation of fermions (i.e. restricting attention to the Hilbert space $H_F = \mathbb{C}^{32}$). An element $a = \{\lambda, q,m\}\in A_F$ can be represented efficiently as  as an $8\times 8$ block diagonal matrix on an element of $\Psi\in H_F$ as~\cite{boyle2019standard}
\begin{align}
L_a \Psi= \left(\begin{tabular}{c c | c c}
\multirow{2}{*}{$q$} & \multirow{2}{*}{}& \multirow{2}{*}{}& \multirow{2}{*}{}\\
& & & \\
\multirow{2}{*}{}& \multirow{2}{*}{$q_\lambda$}& \multirow{2}{*}{}& \multirow{2}{*}{}\\
& & & \\
\hline
\multirow{2}{*}{} & \multirow{2}{*}{}& \multirow{2}{*}{$m$}& \multirow{2}{*}{}\\
& & & \\
\multirow{2}{*}{}& \multirow{2}{*}{}& \multirow{2}{*}{}& \multirow{2}{*}{$\lambda$}\\
& & & \\
\end{tabular}\right)
\left(\begin{array}{c c c c | c c c c}
& & & & {\color{red}u_L} & {\color{green}u_L} & {\color{blue}u_L} & \nu_L\\
& & & & {\color{red}d_L} & {\color{green}d_L} & {\color{blue}d_L} & e_L\\
& & & & {\color{red}u_R} & {\color{green}u_R} & {\color{blue}u_R} & \nu_R\\
& & & & {\color{red}d_R} & {\color{green}d_R} & {\color{blue}d_R} & e_R\\
\hline
{\color{red}\overline{u}_L}& {\color{red}\overline{d}_L} & {\color{red}\overline{u}_R} & {\color{red}\overline{d}_R} & & & &\\
{\color{green}\overline{u}_L} & {\color{green}\overline{d}_L}& {\color{green}\overline{u}_R} & {\color{green}\overline{d}_R}& & & &\\
{\color{blue}\overline{u}_L} & {\color{blue}\overline{d}_L} & {\color{blue}\overline{u}_R} & {\color{blue}\overline{d}_R} & & & &\\
\overline{\nu}_{L} & \overline{e}_{L} & \overline{\nu}_{R} & \overline{e}_{R} & & & &\\
\end{array}\right),\label{mataction}
\end{align}
where $\lambda\in\mathbb{C}$ is a complex number, $q$ is the standard representation of the quaternion $q\in\mathbb{H}$ as a complex $2\times 2$ matrix, and the $2 \times 2$ block $q_\lambda$ is the corresponding diagonal embedding of $\lambda\in \mathbb{C}$ in $\mathbb{H}$.
The element $m$ is given by a $3\times3$ complex matrix.  The left- or right-representation of an element $a\in A_F$ on $h\in H_F$ is given, respectively, by the left matrix product $L_a\Psi =a\Psi$ or the right matrix product $R_a\Psi= \Psi a$. 

For the elements of $\Psi\in H_F$, the fermions (and anti-fermions) of a single standard model generation are expressed as an $8\times 8$ block-off-diagonal matrix, where the 16 components in the upper-right block correspond to the 16 Weyl spinors in a single generation of fermions, while the 16 components in the lower-left block are the corresponding anti-fermions.  Here we have used  red, green, and blue to indicate the three quark colors, and indicate how the three quark columns (or  anti-quark rows) transform into one another like a triplet (or anti-triplet) under strong $SU(3)$.   This representation can also be equipped with a charge conjugation operator $J_F$, which maps particles to anti-particles, and is given simply by Hermitian conjugation: $J_F\Psi = \Psi^\dagger$. The left and right actions of the algebra are furthermore  related by $R_a = J_FL_a^\dagger J_F$.  A grading operator $\gamma_F$ can similarly be constructed, which associates a `$+1$' or `$-1$' to the fermions according to  their usual chirality assignments. We provide an explicit representation of $\gamma_F$ in Appendix~\ref{gamma} for the curious reader, although it will not play a key role in the remaining discussion.

Before discussing the symmetries of the representation given in equation~\eqref{mataction}, or introducing a metric, or constructing dynamics, notice that the noncommutative standard model runs into an awkward problem (see Appendix~\ref{appenA} for how the symmetries of the corresponding particle representations are determined).  The above construction suffers from an over counting in fermionic degrees of freedom by a factor of four~\cite{Lizzi_1997}. The reason for this is that particles and anti-particles, as well as	left and right chiralities are accounted for  both in the internal  Hilbert space $H_F$, as well as in  the `external' Hilbert space of Dirac spinors ${\cal H}_c$ (this must be  done in order to obtain the correct representations under both gauge and local Lorentz symmetries). When the total Hilbert space space is formed by taking the product $H = H_c\otimes H_F$, the  result is an over counting by a factor of four. In order to solve this problem, notice that the total Hilbert space can also be equipped with natural charge conjugation and grading operators ${\cal J} = {\cal J}_c\otimes J_F$ and $\Gamma = \gamma_c\otimes \gamma_F$, where $J_c$ and $\gamma_c$ are respectively the usual charge conjugation and grading on ${\cal H}_c$\footnote{See \cite{Farnsworth_2017,bizi2018thesis} for further information about taking the product between geometries.}. The `fermion quadrupling' problem can then be solved by only considering  those elements of the input Hilbert space $H$ which satisfy the following two requirements~\cite{Barrett_2007}:
\begin{align}
J\Psi&=\Psi, & \Gamma \Psi&=\Psi\label{barretcons}
\end{align}
Or in other words, the over-counting problem is solved by simultaneously imposing both Weyl and Majorana conditions on the spinor fields of the product geometry (see also~\cite{besnard2019uniqueness,Connes_2006} for alternative approaches).

Unfortunately, while projecting out the unwanted spinor fields in this way does result in the correct fermion counting and representations under the gauge  and Lorentz symmetries, the spinor fields that remain are no longer compatible with the  representation of the standard model coordinate algebra. The problem is that the Majorana condition amounts to imposing a `Hermiticity-like' condition on spinors~\cite{boyle2019standard}. Just as the product of two Hermitian matrices is not in general Hermitian, Majorana spinors are not in general closed under the action of associative matrix coordinate algebras.
In particular, the space of Majorana spinors $\{\Psi\in H:J\Psi = \Psi\}$ is  not  closed under the action of the coordinate algebra $A$ on $H$ because the algebra representation does not commute with the charge conjugation operator $[L_f,J]\neq 0$ for $f\in A$.  In general, this will also be true for any noncommutative,  associative algebra, and what this means conceptually is that the  standard model can not correspond directly to any associative, noncommutative geometry. One might of course argue that it is really only the physical fields that must be compatible with the Weyl condition, and not the underlying geometry itself, but this negates the whole point of the geometric construction.

A simple solution to this apparent obstruction  presents itself if one allows for the possibility of nonassociative coordinate algebras. Hermitian matrices are not in general closed under the associative matrix product, but are instead closed under the symmetrized `Jordan' product. That is, while the matrix product $XY$ of two hermitian matrices $X$ and $Y$ is not in general Hermitian, the symmetrized product $\frac{1}{2}(XY + YX)$ will always be. Equipped with such a symmetrized product, the vector space of Hermitian $n\times n$ matrices form a special kind of nonassociative algebra known  as a `Jordan' algebra.   In a similar way,  while the Majorana condition on spinors is not in general compatible with the action of associative, noncommutative coordinate algebras, it is in general compatible with the symmetrized action of the Hermitian elements of a coordinate algebra. This is because the symmetrized action commutes with the charge conjugation on $H$, i.e. $[L_f+R_f,J] = [L_f+J(L_f)^\dagger J,J]= 0$ for all selfadjoint $f\in A$. It is therefore natural to wonder if phenomenologically interesting geometries might exist that are coordinatized by Jordan algebras. Indeed, in~\cite{boyle2019standard} the Jordan coordinate algebra and representation that most closely describes the standard model spinor content and symmetries has been found, along with a natural extension, which describes the Pati-Salam model. In what follows we provide the conceptual justification for considering these geometries seriously, and explicitly construct the first set of tools that will be required for describing the dynamics of realistic models.

\section{Jordan Coordinate algebras}
\label{jordsec}
A number of concerns might arise when considering real Jordan coordinate algebras: (i) First, Connes reconstruction theorem for Riemannian geometries relies on commutative $C^*$-algebras, which are complex. It is therefore natural to wonder  if Jordan geometry is really able to generalize Riemannian geometry, or if too much topological information is lost when  restricting to the selfadjoint (i.e. real) sub-algebra of a $C^*$-algebra. (ii) secondly, Connes provides a formula for calculating distances that coincides with the usual notion of geodesic distance on Riemannian geometries, but which also continues to make sense for finite and discrete noncommutative geometries. Does this  formula still make sense   when restricting to the selfadjoint elements of a noncommutative coordinate algebra? (iii) Finally, do we lose information about the differential structure of the underlying geometry when we restrict to the maximal  selfadjoint sub-algebra of a complex $C^*$-algebra? In this section we explain that each of these concerns is unfounded.

\subsection{Topology}
\label{Section_algebras_topology}

The first  step in describing a geometry comes with  topology. In the familiar setting of Riemannian geometry, the topology of a manifold $M$ is encoded  entirely  by the algebra of complex  functions defined over that manifold $C^\infty(M,\mathbb{C})$. This is the algebra from which the familiar coordinate functions $x^\mu$ are drawn. In particular, a famous theorem by Gelfand, Naimark, and Segal establishes that given a unital, commutative $C^*$-algebra $A$, it is always possible to build a compact space $M$, such that $A$ is interpreted as the algebra of continuous functions defined over $M$~\cite{MARTINETTI_2005,l1997introduction}. 
\begin{align}
\text{Compact topological space $M$} \Longleftrightarrow \text{Commutative, unital $C^*$-algebra  $A$}.\label{GelfandNaimark} 
\end{align}
But how does this correspondence work? Going from left to right is easy. The  complex functions defined over a manifold $M$ form a commutative algebra $A$, with the following pointwise operations:
\begin{subequations}
\begin{align}
(f+g)(x) &= f(x)+g(x),\\
(fg)(x) &= f(x)g(x),
\end{align} 
\label{representcommutative}
\end{subequations}
for $f,g\in A$, $x\in M$. Commutativity arises from the fact that complex numbers commute. This algebra is also equipped with a natural involution\footnote{An involution is an anti-linear map $*:A\rightarrow A$ satisfying the properties $(f^*)^*= f$ and $(fg)^* = g^*f^*$ for all $f,g\in A$. For the algebra of complex $n\times n$ matrices, for example, this is just the familiar conjugate transpose.} and norm
\begin{align}
f^*(x)&=\overline{f(x)},\\
||f|| &= \sup_{x\in M}|f(x)|,\label{normie}
\end{align}
which turn  $A$ into what is known as a complex $C^*$-algebra (i.e. closed in the norm topology and such that $||f||^2 = ||f||||f^*||$). 
Going from right to left in equation \eqref{GelfandNaimark} is less obvious, but comes from the key observation that the points on a manifold $x\in M$ can be thought of as  irreducible  representations of the coordinate algebra $A$. In particular, because $A$ is commutative and associative, all of its irriducible representations will be one dimensional. What  equation~\eqref{representcommutative} is  really saying, then,  is that  the points on a manifold provide complex irreducible representations $\pi_x(f) = f(x)$, satisfying the usual properties
\begin{subequations}
\begin{align}
\pi_x(f+g) &= \pi_x(f) + \pi_x(g),\\
\pi_x(fg) &= \pi_x(f)\pi_x(g),
\end{align}\label{representcommutative2}
\end{subequations}
\hspace{-.11cm}for $f,g\in A$, $x\in M$. The maps $\pi_x:A\rightarrow \mathbb{C}$ satisfying equation~\eqref{representcommutative2} are also known as the characters of $A$. Given an abstract, unital, commutative, complex $C^*$ algebra $A$, its set  of characters $K(A)$ form a compact topological
space, hence the other half of the theorem\footnote{The set $K(A)$ of characters on a unital, commutative $C^*$-algebra is made into a topological space by equipping it with the topology of pointwise convergence on $A$~\cite{l1997introduction}.
}.
The key insight is that a  point $x\in M$ can be seen as an object on which coordinate functions $f\in A$ are evaluated or equivalently, as objects (characters) to be evaluated on functions in order to give numbers (or in other words to provide a one dimensional complex representation).


\subsubsection*{Noncommutative coordinate algebras and topology}

The correspondence given in equation~\eqref{GelfandNaimark} allows one to think either in terms of compact topological spaces, or equivalently in terms of commutative, unital $C^*$-algebras. Once this correspondence is established, however, it is natural to wonder if it might generalize to include algebras that are also noncommutative, and/or nonassociative. This second option has proven less popular (although see~\cite{Boyle_2014,boyle2019standard,farnsworth2013nonassociative,Farnsworth_2015,boyle2016new}), and most attention has focused on the noncommutative case.  In other words, most attention has focused  on whether it is possible, starting with a noncommutative $C^*$-algebra $A$, to construct a space $M$, which is thought of as a noncommutative space over which the elements of $A$ play the role of coordinate functions. What should play  the role of `points' in a `noncommutative' topology? The characters of an algebra (i.e. maps from the algebra to the complex numbers satisfying~\eqref{representcommutative2}) are no longer the appropriate objects to work with because they lose the noncommutative information held in the algebra (because complex numbers commute). Instead, in the noncommutative setting the appropriate tools to extract the topological information are the states of
an algebra. The states of an algebra are linear maps $\pi_x:A\rightarrow \mathbb{C}$, which are positive $\pi_x(f^*f)\geq 0\in \mathbb{R}$, $\forall f\in A$, and which satisfy $\pi_x(\mathbb{I}) = 1$, where $\mathbb{I}$ is the unit of $A$.

The set of states $S(A)$  of a unital $C^*$-algebra $A$ is convex, which means that
any state $\pi_x$ can be decomposed as~\cite{MARTINETTI_2005}
\begin{align}
\pi_x = \lambda \phi + (1 - \lambda)\phi'
\end{align}
where $\phi,\phi'\in S(A)$ and $\lambda\in [0,1]$. The extremal points of $S(A)$, i.e. the states for which the only convex combination is trivial ($\lambda = 1$), are called the pure states of $A$. When the coordinate algebra is commutative its characters and pure states coincide, and so it is natural to think of   pure states as the generalization of `points' in the noncommutative setting. Furthermore, just as in the commutative setting, there is a well understood correspondence between the pure states of a $C^*$-algebra $A$, and the elements of irreducible representations of $A$ (known as the GNS construction~\cite{Falceto_2012}). 

To give a simple (but physically relevant) example, consider the $C^*$-algebra of $2\times 2$ complex matrices $A = M_2(\mathbb{C})$. This algebra has an irreducible representation on the $2$ dimensional complex Hilbert space $H = \mathbb{C}^2$. If $\braket{}{}$ denotes the inner product on $H$, then the map $\rho_\Psi:A\rightarrow \mathbb{C}$ given by
\begin{align}
\rho_\Psi(f)=\bra{\Psi}f\ket{\Psi}
\end{align}
where $f\in A$, and where $\Psi$ is a unit norm vector in $H$, acts as a pure state on $A$. In fact, up to an overall phase, there is a one-to-one correspondence between the  pure states of $A$ and the unit norm elements of $H$. If we introduce an orthonormal  basis $n_i$, $i=1,2$ on $H$, then a general unit norm element can be expressed as:
\begin{align}
\Psi = e^{i\chi}(\cos[\theta/2]n_1 + e^{i\phi}\sin[\theta/2]n_2),
\end{align} 
where $\chi$ is an irrelevant phase, and $0\le \theta\le\pi$, $0\le \phi<2\pi$. This representation is unique except for the case in which $h$ is equal to one of the unit vectors $n_1,n_2$. A useful visualization of the space of pure states is then given in terms of the unit sphere in $\mathbb{R}^3$, where a given state $\vec{\psi}$ is paramaterized by
\begin{align}
\vec{\psi} = (\sin[\theta]\cos[\phi],\sin[\theta]\sin[\phi],\cos[\theta]).
\end{align}
This construction is nothing other than the familiar Bloch Sphere~\cite{Alfsen1998}.

\subsubsection*{Jordan coordinate algebras and topology}

As reviewed in section~\ref{Section_NCGSM}, the space of standard model fermions is not closed under the action of the noncommutative algebra that has been proposed to coordinatize the internal geometry of the standard model. Instead, standard model fermions are only compatible with the `symmetrized' action of the selfadjoint elements of the coordinate algebra. The   selfadjoint elements of a $*$-algebra $A$ are given by the set $A_{sa}=\{f\in A: f^* = f\}$, where $*$ is the involution on $A$. The selfadjoint elements of an associative $*$-algebra are themselves not in general closed under the basic operations of the algebra. In particular, the product $fg$ between two selfadjoint elements $f,g\in A$  is not in general selfadjoint unless $f$ and $g$ commute. Instead, the selfadjoint  elements are closed under the symmetrized  product~\cite{Alfsen1998,Falceto_2012}
\begin{align}
f\circ  g = \frac{1}{2}(fg+gf),\label{JordanprodZ}
\end{align}
where we use juxtaposition to denote the original associative product on $A$. The vector space $A_{sa}$, when equipped with the `symmetrized' product `$\circ$', forms what is known as a real `Jordan' algebra. A real Jordan algebra $A_{sa}$ is a real vector space,  equipped with a bi-linear (abstract) product `$\circ$' that satisfies the following two properties~\cite{Falceto_2012,townsend2016jordan}:
\begin{subequations}
	\begin{align}
	f\circ g &= g\circ f, & &(\text{Commutativity}),\\
	(f\circ g)\circ f^2&= f\circ (g\circ f^2), & &(\text{Jordan Identity}), \label{Jordo1}
	\end{align}\label{JordanId1}
\end{subequations}
\hspace{-.13cm}for $f,g\in A_{sa}$, and where $f^2 = f\circ f$. Note that Jordan algebras are not in general associative, but instead satisfy the weaker `Jordan' identity. For a full classification of the finite-dimensional Jordan algebras see~\cite{Jordan34}, or for a more in depth general introduction see~\cite{Alfsen2003}.

Jordan algebras that can be constructed by `symmetrizing' the product on an associative algebra are known as `special'. In this paper we are primarily interested in special Jordan algebras obtained by  restricting  complex $C^*$-algebras to their self-adjoint elements (similar work focusing on the exceptional Jordan algebra is also currently being pursued~\cite{Dubois_Violette_2016,Carotenuto_2018,Dubois_Violette_2019}). A natural concern, since we are interested in reconstructing geometries, is whether important topological information is lost in the process. Fortunately this turns out not to be the case. The selfadjoint elements of a complex $C^*$-algebra form a special kind of algebra known as a Jordan-Banach-Lie (JBL) algebra, and this subalgebra contains the full topological information of the $C^*$-algebra from which it is formed. Furthermore, a unital JBL-algebra is always Jordan isomorphic to the selfadjoint part of a $C^*$-algebra.

A real Jordan-Lie algebra is a real Jordan algebra $\{A_{sa},\circ\}$, that is additionally equipped with a `Lie' product `$\times$' satisfying~\cite{Falceto_2012}
\begin{subequations}
	\begin{align}
	f_1\times f_2 &= -f_2\times f_1, & &(\text{Anti-commutativity}),\\
	f_1\times (f_2\times f_3)&= (f_1\times f_2)\times f_3+f_2\times (f_1\times f_3), & &(\text{Jacobi Identity}),\label{Jacob}
	\end{align}\label{LieId1}
\hspace{-.13cm}for $f_i\in A_{sa}$. Notice in particular that the Lie product is not in general associative, but instead satisfies  the weaker `Jacobi' identity.
Furthermore the Jordan and Lie products on a Jordan-Lie algebra are compatible in the sense that the following identities are  satisfied:
	\begin{align}
	f_1\times (f_2\circ f_3)&= (f_1\times f_2)\circ f_3+f_2\circ (f_1\times f_3), & &(\text{Leibniz Identity}),\label{leib}\\
	\kappa(f_3\times f_1)\times f_2&=(f_1\circ f_2)\circ f_3-f_1\circ (f_2\circ f_3), & &
	(\text{Associator Identity}),\label{AssociatorIdentity}
	\end{align}\label{JBLderivationidentities}
\end{subequations}
\hspace{-.13cm}for $f_i\in A_{sa}$, and where $\kappa$ is a positive real number. A Jordan-Lie algebra $\{A_{sa},\circ,\times\}$ can be made into a Jordan-Banach-Lie algebra by equipping it with a  norm $||.||$ that satisfies~\cite{Falceto_2012,Alfsen1980,Alfsen1978,Alfsen1998}
\begin{subequations} 
\begin{align}
||f\circ g||&\leq ||f||||g||,\\
||f\circ f||&= ||f||^2,\\
||f\circ f||&\leq ||f\circ f + g\circ g||,
\end{align}
\end{subequations} 
\hspace{-.061cm}for all $f,g\in A_{sa}$.

Starting with a JBL-Algebra $\{A_{sa},\circ,\times,||.||\}$, one is always able to construct from it an associative complex $C^*$-algebra $A$. In particular, an associative product can be defined on the complexification $A = A_{sa}\oplus i A_{sa}$, by making use of both the Jordan and Lie products~\cite{Falceto_2012}
\begin{align}
fg = f\circ g - i \sqrt{\kappa}f\times g\label{Cstarproduct}  .
\end{align}
Associativity follows from the Jacobi identity given in \eqref{Jacob}, and the Leibniz identity given in \eqref{leib}. When further equipped  with the norm $||f + ig||^2 =
||f||^2 + ||g||^2$, the algebra $A$ becomes a $C^*$-algebra whose involution is $(f+ig)^* = f-ig$. Going in the other direction, the real (or selfadjoint) part of $A$ is  precisely $A_{sa}$, and the norm on $A$ induces a norm on $A_{sa}$. In this way one is able to go back and forth between a $C^*$-algebra $A$ and its corresponding Jordan-Banach-Lie algebra $A_{sa}$. 

The key point of interest in this paper is that the state space $S(A)$ of a complex $C^*$-algebra is determined entirely by the self-adjoint part of $A$, which forms a JBL-algebra. The space of states of a real Jordan algebra $A_{sa}$  consists of all those
real linear functionals $\pi_x:A_{sa}\rightarrow \mathbb{R}$ that are positive $\pi_x(f^2)\geq 0$ and normalized $\pi_x(\mathbb{I})=1$~\cite{Alfsen1998,Alfsen1978}. Just as occurs for $C^*$-algebras, the  states of a JBL-algebra form a convex set. There is a natural identification between the states $S(A_{as})$ of $A_{sa}$ and the states $S(A)$ of the corresponding $C^*$-algebra $A$~\cite{Falceto_2012}. In particular, given a state $\pi_x$ of $A_{as}$ a linear functional $\overline{\pi}_x$ can be defined on $A$ by extending $\pi_x$ by linearity: $\overline{\pi}_x(f+ig) = {\pi}_x(f)+i{\pi}_x(g)$. The converse is trivial as $ {\pi}_x(f)=\overline{\pi}_x(f)$ for all elements $f\in A_{sa}$. In short, nothing is lost, topologically speaking, when shifting attention from $C^*$-algebras, to JBL-algebras. The full topological data of a $C^*$-algebra is encoded in its maximal JBL-subalgebra.

\subsection{Measuring distances}
\label{secjorddist}
Geometry is more than just  points on a manifold. At   a  minimum we would also  like to  be able to define a notion of distance between points. Furthermore,  because we are interested in generalized notions of geometry, for which   it is easier to deal with coordinate algebras than it is to deal with the topological spaces they define,    it would be useful to have a definition of distance that only needs to make reference to coordinate algebras. Fortunately an appropriate  definition  has already been found~\cite{Connes1994,Connes2007}, which coincides with the familiar geodesic distance on Riemannian manifolds, but which makes no reference to any underlying manifold, and which also continues to make sense in the noncommutative and discrete settings. The question, one might ask, is whether this standard definition of distance continues to make sense in the Jordan setting.

Consider a one dimensional Riemannian manifold $M=\mathbb{R}$ coordinatized by $A= C^\infty(\mathbb{R},\mathbb{C})$ (i.e. the algebra of smooth complex coordinate functions defined over the real line). Without reference  to the manifold itself, one way that the distance between two points `$x$' and `$y$' on $\mathbb{R}$ is able to be defined,  is  as the excursion of a carefully  chosen function $f\in A$ between the two points. In particular, if we select $f$ to be the function with the maximum possible excursion subject to the condition that its derivative is never greater than one, then this recovers precisely the usual geodesic distance $|x-y|$ between the points $x,y\in\mathbb{R}$. In other words we can define the distance as:
\begin{align} 
d(x,y) = \sup_{f\in C^\infty(\mathbb{R})}\{|f(x)-f(y)| : \sup_{z\in\mathbb{R}}|f'(z)|\leq 1\}.\label{distancecon}
\end{align} 
This definition  relies not only on the coordinate algebra, but also on knowing what a derivative is. For higher dimensional manifolds the derivative is readily replaced with a gradient, but  a notion of derivation is needed that also continues to make sense  in a much more general setting. Sticking with our simple example for the time being, notice that if  the coordinate algebra $A= C^\infty(\mathbb{R},\mathbb{C})$ is represented on the Hilbert space of square intergrable spinors $H=L^2(\mathbb{R},S)$, then this allows us to represent the derivative of a function in terms of the Dirac operator acting on $H$.  
\begin{align}
f'(x)\Psi(x) = \frac{df(x)}{dx}\Psi(x) = \frac{d}{dx}f(x)\Psi(x) - f(x)\frac{d}{dx}\Psi(x) = [D,f(x)]\Psi(x),
\end{align}
where $\Psi\in H$ and $D$ is the derivative along the manifold $d/dx$.  Making use of  equation \eqref{normie}, we have $||[D,\pi(f)]|| = \sup_{x\in \mathbb{R}}|f'(x)|$, which allows us to  re-express equation~\eqref{distancecon} as
\begin{align}
d(x,y) = \sup_{f\in A}\{|\pi_x(f)-\pi_y(f)|:||[D,f]||\leq 1\}\label{connesdistance}.
\end{align}
This is Connes' distance formula~\cite{Connes2007}, and it continues to make sense for   higher dimensional Riemannian manifolds, as well as for  spaces coordinatized by noncommutative and discrete algebras. The main extra ingredient is the notion of a generalized Dirac operator that satisfies some appropriate properties~\cite{Connes1994}.  Connes' notion of distance becomes especially interesting in situations where the usual classical definition of the distance as the length of the shortest path between two points is no longer available. 
In this paper we are interested in restricting to the maximal JBL sub-algebras of  noncommutative $C^*$-algebras, and in this case it is natural to wonder whether Connes notion of distance continues to make sense. 

For a geometry coordinatized by a noncommutative $C^*$-algebra, consider a function $f$ that reaches the supremum in~\ref{connesdistance}, such that $||[D,f]||\le 1$, and  $|(\pi_x - \pi_y)(f)| = \mathrm{dist}(x,y)$, and let  $\theta:= \mathrm{Arg}((\pi_x-\pi_y)(f))$. In this case the supremum is also reached by the selfadjoint element $g = 1/2(fe^\theta + f^* e^{-\theta})\in A_{sa}$ because~\cite{Iochum_2001}
\begin{align}
||[D,g]||&\le \frac{||[D,f]||}{2} +\frac{||[D,f]||}{2}\le 1, \\
|(\pi_x-\pi_y)(g)| &= |\frac{\mathrm{dist}(x,y)}{2} + \frac{\overline{\mathrm{dist}(x,y)}}{2}| = \mathrm{dist}(x,y).
\end{align} 
In practice this means that we can actually restrict attention in Connes distance formula to those elements of the coordinate algebra that are selfadjoint. Not only that, but  the supremum is also met for the element $k = g + ||g||\mathbb{I}\in A^+$, and so we can really restrict attention further to the positive elements $A^+_{sa}$ of $A$ (i.e. the self adjoint elements with eigenvalues greater than or equal to zero).

Now suppose that $||[D,k]||< 1$, and take $l :=k/||[D,k]||\in A^+_{sa}$, then $||[D,l]||=1$, and
\begin{align}
 |\pi_x(l)-\pi_y(l)| &= \frac{\pi_x(k) - \pi_y(k)}{||[D,k]||}>|\pi_x(k) - \pi_y(k)|, 
\end{align}
which is impossible because  $k$ was chosen to reach the supremum. We therefore have that  $||[D,k]||= 1$ when Connes' distance formula reaches a supremum, and so equation~\eqref{connesdistance} can be re-expressed as~\cite{Iochum_2001}: 
\begin{align}
d(x,y) = \sup_{f\in A^+_{sa}}\{|\pi_x(f)-\pi_y(f)|:||[D,f]||= 1\}\label{connesdistance2}.
\end{align}

The distances of a noncommutative geometry are `observable' quantities. When constructing gauge theories, for example, the discrete, internal, `flat', Dirac operator $D_F$ corresponds (roughly speaking) to the mass-matrix of the theory. The distances measured on the internal space are therefore ultimately related to the mass spectrum of the theory~\cite{Martinetti:2001zz}. The take home message is that when determining `observables', it is the selfadjoint elements of a noncommutative coordinate algebra that are physically relevant. In other words, with regards to both topology and distance measurements, the physically relevant data of a noncommutative, complex $C^*$-algebra appears to be entirely contained within its maximal JBL-subalgebra.

\subsection{General coordinate transformations}
\label{sectjordsym}
A common theme in theories with small extra dimensions is the unification of the four known fundamental forces, with the idea being that the  local gauge symmetries are contained within the diffeomorphism group of the higher dimensional  total space. This same idea continues to apply when considering geometries in which the internal space is finite and discrete. The symmetries of a geometry correspond, roughly speaking, to the automorphisms of its coordinate algebra\footnote{The symmetries of a geometry correspond more correctly  to the automorphisms of the representation of the coordinate algebra $A$ on the Hilbert space of spinors $H$~\cite{Farnsworth_2015}. In general these symmetries will be slightly larger than the automorphism group of the coordinate algebra $A$, but we will not concern ourselves with this subtlety here. This is already a familiar feature in Riemannian spin geometry, in which the symmetries include not only the diffeomorphisms of the manifold, but also to local rotations in the spin bundle.}. In this section, we explain how the symmetries of a noncommutative coordinate algebra are exactly the same as those defined on its maximal JBL sub-algebra. Intuitively this makes sense, as the symmetries of an algebra should not map selfadjoint elements to those that are anti-selfadjoint (this is in fact a defining property as we outline below in equation~\eqref{Def_Auto_2}).
 
The symmetries of an algebra are given by its automorphisms. An automorphism $\alpha$ is an invertible linear map from an algebra to itself, which preserves the structure of the algebra. That is, it respects the product on $A$: 
\begin{subequations}
\begin{align}
\alpha(fg) &= \alpha(f)\alpha(g), \label{Def_Auto_Lieb}
\end{align}
for all $f,g\in A$. If $A$ is a $*$-algebra, then those automorphisms which also respect the involution on $A$ are called $*$-automorphisms: 
\begin{align} 
\alpha(f^*) &= \alpha(f)^*,  \label{Def_Auto_2}
\end{align}
\end{subequations}
for all $f\in A$.  Of particular interest in this paper  are those automorphisms that are continuously connected to the identity $\mathbb{I}$. An automorphism $\alpha$, which is infinitessimally close to the identity map can then be written as $\alpha = \mathbb{I} + \delta$ where $\delta$ is a derivation element satisfying:
\begin{subequations} 
\begin{align}
\delta(fg) &= \delta(f)g+f\delta(g), \label{Def_Der_1}
\end{align}
for all $f,g\in A$. If $A$ is also a $*$-algebra, then a $*$-derivation $\delta$, is a derivation which respects the involution on $A$:
\begin{align} 
\delta(f^*) &= \delta(f)^*, \label{Def_Der_2}
\end{align}
\label{derivationidentities}
\end{subequations}
\hspace{-0.12cm}for all $f\in A$. The derivations of $A$ are the infinitesimal generators of the automorphisms of $A$; they form a Lie algebra, with Lie product given by
$\delta_1\times  \delta_2 = \delta_1  \delta_2 - \delta_2  \delta_1$ (where juxtaposition on the right hand side denotes composition of operators on $A$). 

\subsection*{Riemannian geometry}
Consider the coordinate algebra $A=C^\infty(M,\mathbb{C})$ defined over a Riemannian manifold $M$. The diffeomorphisms on $M$ are in one-to-one correspondence with the coordinate transformations of $C^\infty(M,\mathbb{C})$. In particular, for any diffeomorphism $\phi: M \rightarrow M$, one can construct a map  $\alpha_\phi f(x) = f(\phi^{-1}x)$, where $ f(x)\in C^\infty(M,\mathbb{C})$. The maps $\alpha_\phi$ are known as `outer automorphisms' and satisfy all the properties required of $*$-automorphisms. In particular they preserve the structure of the product and involution on $A$:
\begin{subequations} 
\begin{align}
\alpha_\phi (f_0f_1)(x) &= (f_0f_1)(\phi^{-1}x)\nonumber\\
&= f_0(\phi^{-1}x)f_1(\phi^{-1}x)\nonumber\\
&= \alpha_\phi f_0(x)\alpha_\phi f_1(x) = (\alpha_\phi f_0 \alpha_\phi f_1)(x),\label{eq_aut_prod}\\
\alpha_{\phi}(f_0^*)(x)&= f_0^*(\phi^{-1}x)\nonumber\\
&= \overline{f_0(\phi^{-1}x)}\nonumber\\
&= \overline{\alpha_\phi f_0(x)} = \alpha_{\phi}(f_0)^*(x) 
\end{align}
\end{subequations} 
\hspace{-.06cm}where $f_i\in C^\infty(M,\mathbb{C})$. If we further consider those automorphisms that are infinitesimally close to the identity, we find that their generating derivations take the form $\delta_V(f_0) = V^\mu\partial_\mu f_0$, where the $V(x)^\mu$ are real valued coefficients. Such derivations clearly satisfy the properties of $*$-derivations given in equations \eqref{derivationidentities}. In addition, they  form a Lie algebra (of vector fields)  with the Lie product given by $\delta_V\times \delta_W = (V^\nu (\partial_\nu W^\mu)- W^\nu (\partial_\nu V^\mu ))\partial_\mu$.

\subsubsection*{Noncommutative coordinate algebras and coordinate transformation}

Noncommutative algebras generally have additional automorphism known as `inner automorphisms', which are constructed from elements of the algebra itself.  In particular, given  a unital, associative $*$-algebra $A$, we can define the map $\alpha_u:A\rightarrow A$
  \begin{align}
  \alpha_u f = ufu^{-1},\label{assocunit} 
  \end{align}
  where $f,u\in A$.    It is easy to check that such maps satisfy the automorphism property
  \begin{subequations} 
  \begin{align} 
  \alpha_u (fg) &= ufgu^{-1}\nonumber\\
  &= ufu^{-1}ugu^{-1} = \alpha_u(f)\alpha_u(g),
  \end{align} 
  for $f,g\in A$. Furthermore, if $u$ is a unitary element in $A$ (i.e. satisfying $u^* = u^{-1}$), it is easy to check that such maps $\alpha_u$ act as $*$-automorphisms
  \begin{align}
\alpha_u (f^*)& = uf^*u^*\nonumber\\
&=(ufu^*)^* = \alpha_u (f)^*. 
	\end{align}
	\end{subequations} 
Note that the unitary elements $\{u\in A:u^*=u^{-1}\}$ are generated by anti-hermitian elements $\{a\in A:a^* = -a\}$ through exponentiation, which allows us to express inner automorphisms given in \eqref{assocunit} as
	\begin{align}
	\alpha_uf = e^{a}fe^{-a} = e^{L_a - R_a}f = e^{\delta_a}f,\label{assocderiv}
	\end{align}
	for where we have defined $\delta_a = L_a - R_a = [a,\_]$, and where we are using the standard `left-right' notation $L_af = af$, and $R_af = fa$~\cite{Schafer1966}. The elements $\delta_a$ act as $*$-derivations on $A$, and form a Lie algebra when equipped with the Lie product $\delta_a\times \delta_b = \delta_a\delta_b-\delta_b\delta_a $. 
	
	As an example, if we take our coordinate algebra to be the $C^*$-algebra of $n\times n$ complex matrices $A = M_n(\mathbb{C})$, then all $*$-derivations are `inner' and take the form $\delta_a = (L_a - R_a)$, where  $a = -a^*\in A$.	The Lie algebra of inner $*$-derivations is given by $su(n)$, and generates the Lie group of inner $*$-automorphisms  $SU(n)/\mathbb{Z}_n$ through exponentiation. If, further, we take the product between a canonical `Riemannian' geometry coordinatized by $A_c = C^\infty(M,\mathbb{C})$ and a finite internal  geometry coordinatized by $A_F = M_n(\mathbb{C})$, then the product coordinate algebra will be given by $A = C^\infty(M,\mathbb{C})\otimes_{\mathbb{C}} M_n(\mathbb{C}) = C^\infty(M, M_n(\mathbb{C}))$. In this case the $*$-automorphism group will be given by $Diff(M)\ltimes SU(n)/\mathbb{Z}_n$~\cite{kadison1967}, such that  the product $C^*$-algebra  can be understood as encoding the  topological data of an $SU(n)$ gauge theory coupled to Einstein-Hilbert gravity.

\subsubsection*{Jordan coordinate algebras and coordinate transformations}

Jordan algebras in general also have inner automorphisms. In this case, however, they are generated through exponentiation by derivation elements of the form~\cite{Schafer1966}
\begin{align}
\delta_{ab} = [L_a,L_b], \label{jorderiv}
\end{align}
for $a,b\in A$. Notice that, due to commutativity, these derivations can be expressed in terms of  associators $\delta_{ab} = [b,\_,a]$ on $A$. In other words $C^*$-algebras are associative, possibly noncommutative algebras with inner derivations given by commutators, while Jordan algebras are possibly nonassociative, commutative algebras with inner derivations given by associators.

	As an example, if we take our coordinate algebra to be the JBL-algebra of $n\times n$ complex, Hermitian matrices $A = H_n(\mathbb{C})$, then all  derivations are `inner' and take the form $\delta_{ab} = [L_a,L_b]$ $a,b\in A$.	The Lie algebra of inner derivations is given by $su(n)$, and generates the Lie group of inner automorphisms  $SU(n)/\mathbb{Z}_n$ through exponentiation. If, further, we take the product between a canonical `Riemannian' geometry coordinatized by $A_c = C^\infty(M,\mathbb{R})$ and a finite internal geometry coordinatized by $A_F = H_n(\mathbb{C})$, then the product coordinate algebra will be given by $A = C^\infty(M,\mathbb{R})\otimes_{\mathbb{R}} H_n(\mathbb{C}) = C^\infty(M, H_n(\mathbb{C}))$. In this case the automorphism group will be given by $Diff(M)\ltimes SU(n)/\mathbb{Z}_n$, such that the  product Jordan algebra  can be understood as encoding the  topological data of an $SU(n)$ gauge theory coupled to Einstein-Hilbert gravity. In other words, nothing is lost in restricting to the maximal JBL-subalgebra of a noncommutative coordinate algebra.

\subsection{Differential structure}
\label{diffstructure}
Coordinate algebras hold  much more information than just the topological data of a geometry. They also encode information about the \emph{differential} structure. Consider, for instance, a \emph{smooth} Riemannian manifold $M$. In  order to reconstruct $M$ from its coordinate functions, the algebra of \emph{smooth} coordinate functions $A = C^\infty(M,\mathbb{R})$ is required (considered as an abstract  JBL-algebra). If one instead only had access to the  algebra of continuous functions $C^0(M,\mathbb{R})$, then one would lose all of the smooth  structure of the manifold. The key feature that distinguishes the coordinate algebra of smooth functions $A = C^\infty(M,\mathbb{R})$ from the algebra of continuous functions $C^0(M,\mathbb{R})$, is that it is equipped with an algebra of many well defined derivations, namely the vector fields $\delta_V = V^\mu\partial_\mu$ defined over the manifold $M$~\cite{Dubois_Violette_1990}. For Riemannian manifolds the lie algebra of vector fields coincides with the lie algebra of derivations $\mathrm{Der}(C^\infty(M,\mathbb{R}))$ on $C^\infty(M,\mathbb{R})$, and it is this interpretation that generalizes most naturally to product geometries with internal structure.

 Following previous work\cite{Dubois_Violette_2016,Carotenuto_2018}, we take the view that the appropriate notion of a vector field is that of a derivation, and that the analogue of the differentiable structure is encoded in the lie algebra of derivations $\mathrm{Der}(A)$ defined over the coordinate algebra $A$\cite{Dubois_Violette_1990}. A natural concern when restricting attention to the Hermitian sub-algebra of a complex $C^*$-algebra is whether one looses information about the differential structure of the corresponding geometry when doing so. For the coordinate algebras of interest, however, this turns out not to be the case. Consider for example a Riemannian geometry $(M_c,g)$. As discussed in the previous  section, the Lie algebra of $*$-derivations defined over the complex coordinate algebra $C^\infty(M_c,\mathbb{C})$ (i.e. the Lie algebra of vector fields with real coefficients), clearly coincides with the Lie alegbra of derivations defined over the real coordinate algebra $C^\infty(M_c,\mathbb{R})$. The story is similar when considering coordinate algebras with inner derivations. In particular,  the associator identity given in equation \eqref{AssociatorIdentity} relates derivations of a Jordan algebra $\mathrm{Der}(A_{sa})$ expressed in terms of associators, with $*$-derivations of the corresponding  $C^*$-algebra $\mathrm{Der}(A)$ written in terms of the Lie product. Indeed, as shown for the case of most physical interest in the last subsection, the Lie algebra of $*$-derivation defined over the $C^*$-algebra of $n\times n$ complex matrices $M_n(\mathbb{C})$ coincides with the Lie algebra of derivations defined over the JBL-algebra of $n\times n$ complex Hermitian matrices, both of which are given by $su(n)$. In other words, it appears as if all of the physically interesting data of an associative $C^*$-algebra $A$, including differential structure, is really captured by its maximal JBL-algebra $A_{sa}$. 

\subsubsection*{Differential calculi}
\label{NAGPT_Forms}

Following~\cite{Dubois_Violette_2016,Carotenuto_2018},  we take the view that  derivations are the natural generalization of vector fields, and that the natural analogue of differentiable structure for a geometry coordinatized by a Jordan algebra $A_{sa}$,  is encoded in  the lie algebra of derivations $\mathrm{Der}(A_{sa})$. This point of view implies a correspondingly natural generalization of the notion of differential forms in the Jordan setting. In this subsection we present an explicit construction of the derivation based differential graded algebra of forms defined over Jordan algebras of $n\times n$ complex Hermitian matrices.

Consider the derivation algebra $\mathrm{Der}(A_{sa})$ defined over a unital JBL-algebra $A_{sa}$. Given a function $f\in A_{sa}$, and a vector field $\delta_V = V^i\delta_i\in \mathrm{Der}(A_{sa})$, where the $\delta_i$ form a linearly independent basis of derivations\footnote{Note here that the subscript `$i$' on the basis element $\delta_i\in \mathrm{Der}(A_{sa})$ indexes the basis, and is not an element of the coordinate algebra. This should not be confused for example with the double subscript on the  derivation elements given in equation~\eqref{jorderiv} or the single subscript given on derivation element just below equation~\eqref{assocderiv}, which really do correspond to elements of the coordinate algebra (for Jordan and associative coordinate algebras respectively). In this section we are speaking about derivations in a general and abstract manner, and so it is convenient  to index a basis of derivations in the familiar way.}, the object $\delta_V f\in A_{sa}$ can be viewed as a linear map in two distinct ways. To start with, of course, we can view $\delta_V f$ as the action of the linear operator $\delta_V$ on the function $f$. Notice however, that because derivations form a vector space, the object $\delta_V f$ is also linear in the argument `$V$', in the sense that 
\begin{align}
\delta_{V+W} f = \delta_{V} f+ \delta_{W} f,  
\end{align}
for $\delta_V,\delta_W\in  \mathrm{Der}(A_{sa})$. One can therefore view the object $\delta_Vf$ as a linear operation being performed on the vector field $\delta_V$. This map is denoted by $df:\mathrm{Der}(A_{sa})\rightarrow A_{sa}$, where:
\begin{align}
df(\delta_V) \equiv \delta_V f,
\end{align}
and where `$d$' is known as the  `exterior derivative' on functions. The object $df$ is known as an `exact' one-form, and is completely defined by its action on vector fields. In particular, because each vector field can be expressed in terms of an  orthonormal basis $\delta_V= V^i\delta_i$, we are able to write:
\begin{align}
df = (\delta_if) E^i,
\end{align}
where the $E^i$ satisfy $E^i(\delta_j) = \hat{\delta}^i_j$, and where  $\hat{\delta}$ denotes the Kronecker delta. If the $\delta_i$ span a $k$-dimensional basis of $\mathrm{Der}(A_{sa})$, then the $E^i$ span a $k$-dimensional basis of dual vectors which we denote by $\mathrm{Der}^*(A_{sa})$. The space of one forms is then defined as the free module $\Omega^1 A_{sa} = A_{sa}\otimes \mathrm{Der}^*(A_{sa})$, such that an arbitrary one form $\omega\in \Omega^1 A$ can be expressed as:
\begin{align}
\omega = \omega_i E^i,
\end{align}
for $\omega_i\in A_{sa}$.

The notion of one forms as linear maps from $\mathrm{Der}(A_{sa})$ to $A_{sa}$, can  be generalized to that of `$n$-forms', which are totally antisymmetric, multi-linear maps from $\mathrm{Der}(A_{sa})$ to $A_{sa}$. Given two one forms $\omega=\omega_iE^i$ and $\omega'=\omega_j'E^j$ for example,  a new `two-form' can be defined by taking the generalized `wedge' product:
\begin{align}
\omega\wedge\omega' \equiv \omega_{i}\omega_{j}'(E^i\otimes E^j- E^j\otimes E^i).
\end{align} 
General $n$-forms can  similarly be defined by taking successive wedge products between forms:
\begin{align}
\omega = \omega_{1,...,n}~E^1\wedge...\wedge E^n, \label{JordanWedge} 
\end{align}
with $\omega_{1,...,n}\in A_{sa}$. We will almost always drop the `wedge' sign when taking the product between two or more different forms (i.e. we will usually write $\omega\omega'$ to indicate the wedge product $\omega\wedge\omega'$). Under the product given in Eq.~\eqref{JordanWedge}, the vector space $\Omega A_{sa} = \oplus_n \Omega^n A$, where $\Omega^n A_{sa}$ is the space of $n$-forms with $\Omega^0 A_{sa} = A_{sa}$, forms a graded Jordan algebra that satisfies the following  identities
\begin{subequations}
	\begin{align}
	ab -(-1)^{|a||b|}ba&= 0, & & \text{(Graded Commutativity)}\label{jords10}\\
	\sum_{\{a,b,c\}}(-1)^{|a||c|}(L_aL_{bc}-L_{ab}L_c)&=0,& & \text{(Graded Jordan Identity)}\label{jords11}
	\end{align}
where in the graded Jordan identity we are summing over all even permutations of $a,b,c\in \Omega A_{sa}$, and where we denote the `grading' or the `order' of an element $a\in \Omega A_{sa}$ by $|a|$, i.e. $|a| = m$ when $a\in \Omega^mA_{sa}$. Notice in particular that at order zero these identities reproduce the usual properties of a Jordan algebra of order zero given in equation~\eqref{JordanId1}. In particular, the more familiar form of the Jordan identity given in equation~\eqref{Jordo1}  is easily recovered at order zero by setting $a=b=c$ in equation~\eqref{jords11} (deriving \eqref{jords11} from \eqref{Jordo1} at order zero is only slightly more involved~\cite{Schafer1966,Kac77}).

 A graded Jordan algebra algebra $\Omega A_{sa}$ can be further elevated to a differential graded algebra of forms, by extending the  action of the exterior derivative  `$d$' to the whole of $A$ such that it satisfies the following two properties:
\begin{align}
d^2 &=0, & &\text{(Nilpotency condition)}, \label{DJord00}\\
d(ab) &= d(a)b + (-1)^{|a|}a d(b), & &\text{(Graded Leibniz)} \label{DJord01}
\end{align}  \label{DJord0}
\end{subequations}
for $a,b\in \Omega A_{sa}$.

\textbf{Example: the `Canonical' differential graded algebra of forms.} 

Consider  the canonical (Jordan) coordinate algebra $A_{sa} = C^\infty(M,\mathbb{R})$ corresponding to a Riemannian manifold $M$, with a local basis of coordinate functions  $x^\mu\in A_{sa}$. Any vector field $\delta_V\in \mathrm{Der}(A_{sa})$ can be expanded in terms of this basis $\delta_V= V^\mu\frac{\partial}{\partial x^\mu}$.  Similarly, a dual  basis of forms can  be constructed $E^\mu=dx^\mu \in \Omega^1 A_{sa}$, with the action on vector fields given by $dx^\mu(\delta_V) = V^\nu\partial x^\mu/\partial x^\nu = V^\mu$.  A general one form $\omega\in \Omega^1A_{sa}$  is then given by: 
\begin{align}
\omega =\omega_\mu~dx^\mu, 
\end{align}
where $\omega_\mu\in A_{sa}$. In particular we can  write $df = \partial_\mu f dx^\mu $. Similarly, general $n$-forms are defined by taking successive wedge products between forms:
\begin{align}
\omega = \omega_{1,...,n}~dx^1\wedge...\wedge dx^n,
\end{align}
with $\omega_{1,...,n}\in A_{sa}$. Under this product the algebra $\Omega A_{sa} = \oplus_n \Omega^n A_{sa} $ is both graded commutative and associative, which means that it trivially satisfies the conditions of a graded Jordan algebra as given in equations \eqref{jords10} and  \eqref{jords11}. Furthermore,  $\Omega A_{sa}$ can be elevated naturally to a differential graded algebra of forms, by extending the  action of the exterior derivative `$d$' to arbitrary forms such that it satisfies equations \eqref{DJord00} and \eqref{DJord01}. Notice in particular, that the nilpotency of the exterior derivative on forms $d^2=0$, is automatically ensured by the graded commutativity of the wedge product, the graded Leibniz rule, as well as the commutativity of partial derivatives.


\textbf{Example: The finite coordinate algebra $A_F =H_n(\mathbb{C}$)}

Next, we explicitly construct the derivation based differential graded algebra of forms defined over the JBL-algebra $A_{sa} = H_n(\mathbb{C})$ (i.e. the Jordan algebra of $n\times n$ complex Hermitian matrices). The corresponding $C^*$-algebra $A = A_{sa}\oplus iA_{sa}$ is given by the matrix algebra of $n\times n$ complex matrices.  Given a basis of linearly independent    traceless, complex, Hermitian  matrices $\lambda_i\in A$, $i=1,...,n^2-1$, the associative product on $A$ (see equation~\eqref{Cstarproduct}) can be expressed as:
\begin{align}
\lambda_i\lambda_j = \frac{1}{2}(\underbrace{\frac{1}{n}\hat{\delta}_{ij}\lambda_0 + d_{ijk}\lambda_k}_{\text{Jordan}} + \underbrace{if_{ijk}\lambda_k}_{\text{Lie}}),\label{matprod}
\end{align}
where $\lambda_0 =\mathbb{I}_n$ is the identity element in $A_{sa}$. The  $d_{ijk}$ are real valued and completely symmetric `Jordan' structure constants which define the Jordan  product `$\circ$', while the $f_{ijk}$ are the completely anti-symmetric structure constants that define the Lie  product `$\times$' (i.e. the structure constants of the $su(n)$ Lie algebra). Notice that for this class of Jordan algebras specifically $A_{sa} = H_n(\mathbb{C})$, the dimension of the derivation algebra $\mathrm{Der}(A_{sa}) = su(n)$ coincides exactly with the number of linearly independent traceless Hermitian basis elements $\lambda_i\in A_{sa}$. This allows us to define a particularly elegant basis of $n^2-1$ anti-hermitian inner derivation elements, expressed in terms of the `Lie' algebra structure constants:
\begin{align}
\delta_i = \frac{4}{n}f_{i}^{\phantom{i}jk}[L_{\lambda_j},L_{\lambda_k}],\label{Jordder}
\end{align}
where summation is implied over repeated indices, and where	 $L_{\lambda_{i}}f = \frac{1}{2}(\lambda_i f+f \lambda_i) = \lambda_i\circ f$ denotes the left action under the Jordan product (note that the only distinction between lowered and raised indices is stylistic). Making use of Eq.~\eqref{matprod} it is easy to show that the action of a derivation $\delta_i\in \mathrm{Der}(A_{sa})$ on elements $\lambda_j\in A_{sa}$, is given by:
\begin{align}
\delta_{i}\lambda_j &= -f_{ij}^{\phantom{ij}k}\lambda_k,\label{derjoract}
\end{align}
where the standard normalization  $f^{ijk}f_{ijl}=n\hat{\delta}_{l}^k$ is being used. Notice in particular, that  equation \eqref{derjoract} is really only restating the fact that the coordinate algebra $A_{sa}$ is a JBL-algebra with Lie product
\begin{align}
\lambda_i\times\lambda_j = \delta_{i}\lambda_j
\end{align}
for $\lambda_i\in A_{sa}, i=1,...,n^2-1$.

Having expressed the derivation elements in a nice basis, our next task is to find a similarly nice dual basis of one forms $E^i\in \Omega^1 A_{sa}$, such that $E^i[\delta_j] = \hat{\delta}_{j}^i$. A little work shows that the dual elements take the form:
\begin{align}
E^i  = \frac{16}{n}f^{jki}(\lambda_k\circ \lambda_l)\circ(\lambda_j\circ d\lambda_l),\label{dualbasisJ}
\end{align}
where `$d$' is the order one exterior derivative. The proof that the $E^i$ form a dual basis of one forms is simple but long, and so we provide it separately in  appendix~\ref{AppendB}. Having constructed an explicit basis of derivations and dual forms, an arbitrary one form $\omega\in \Omega^1 A_{sa}$ can be expressed as:
\begin{align}
\omega = \omega_i~ E^i,
\end{align}
where $\omega_i\in A_{sa}$. In particular the action of the exterior derivative is given by $df = \delta_if E^i $, for  $f\in A_{sa}$. Furthermore, general $n$-forms are defined by taking successive wedge products between forms. Given two elements $a =a^{1,...,n}~E_1\wedge...\wedge E_{n} \in \Omega^nA$ and $b= b^{1,...,m}~E_1\wedge...\wedge E_{m}\in \Omega^mA$, their graded Jordan product is given by:
\begin{align}
ab& =(a_{1,...,n}b_{1+n,...,m+n})~E^1\wedge...\wedge E^{n+m}\in A^{m+n}. 
\end{align}
It is simple to  check that the graded commutativity and graded Jordan identities given in equations~\eqref{jords10} and \eqref{jords11} are satisfied by $\Omega A_{sa} = \oplus_n \Omega^n A$. Next, the graded algebra  $\Omega A$ can be promoted to a differential graded algebra of forms by extending the action of the exterior derivative `$d$' to higher order forms such that it satisfies the properties  given in equations~\eqref{DJord00} and \eqref{DJord01}. Following a fairly long, but straightforward calculation (which we provide in appendix~\ref{AppendB}) we find that the action of the exterior derivative on the one form basis is given by
\begin{align}
	dE^i &= \frac{1}{2}f_{\phantom{i}jk}^i E^j\wedge E^k.
\end{align}

Notice, that the differential graded algebra $\Omega A_{sa}$ has a number of properties that distinguish it, and make it much nicer than the  differential graded algebra of forms $\Omega A$ defined over the corresponding noncommutative $C^*$-algebra of $n\times n $ complex matrices $A= A_{sa}\oplus i A_{sa}$~\cite{l1997introduction}. In particular, once one  demands that the exterior derivative satisfy the  nilpotency condition given in equation~\eqref{DJord00}, this automatically implies the graded commutativity, graded Leibniz, and graded Jordan identities given in equations~\eqref{DJord0}. This is very much in line with the story as it occurs in the Riemannian setting, and is in stark contrast to the noncommutative setting, in which the construction is much more  awkward and appears somewhat contrived.

\section{Conclusion and future outlook}
\label{seccon}
Jordan algebras were first introduced in an effort to restructure quantum mechanics purely in terms of physical observables (i.e. the things that can actually be seen in  experiment)~\cite{Jordan33,townsend2016jordan}. In the current paper we have similarly been drawn on physical grounds to consider geometries coordinatized by Jordan algebras. The usual assumption in the literature has been that if the patterns and features of the standard model derive from the underlying structure of a discrete internal geometry, then this geometry should be coordinatized by a noncommutative $C^*$-algebra. The noncommutative geometry that most closely captures the details of  the standard model, while extraordinarily beautiful and insightful, runs into a variety of  problems, and relies  on a number of  nongeometric input assumptions in order to work. In particular, it appears as though the   space of standard model fermions is not compatible with the full noncommutative coordinate algebra that must been selected  in order to accommodate the standard model symmetries (this remains  true for extensions such as the Pati-Salam model~\cite{Chamseddine_2013}). Instead, the standard model fermions are only closed under the symmetrized action of the selfadjoint elements of the noncommutative coordinate algebra~\cite{boyle2019standard}. These elements form a Jordan algebra, which motivates the search for a `Jordan geometry', which is able to accommodate the full details of the standard model.

In this paper we explain that the most important physical data captured by a complex  coordinate $C^*$-algebra appears to be entirely contained within its maximal JBL-subalgebra. Our program is  an attempt to restructure the geometric description of nature purely in terms of physically relevant quantities, and furthermore to make important geometric inferences based purely on  physical observables. Secondly, this paper presents an explicit construction of the differential calculus for the finite dimensional coordinate algebras that appear most important for constructing realistic particle theories (i.e. the Jordan algebras of $n\times n$ complex Hermitian matrices $H_n(\mathbb{C})$). This is the first of a number of tools that will be required for describing the dynamics of physical theories. Constructing the `geometry' of a physical theory consists roughly of two parts: (i) building the underlying geometry itself (for a gravity theory this would be the underlying Riemannian manifold and metric data), and (ii) describing the dynamics of the geometry (for a gravity theory this might come from the familiar Einstein-Hilbert action, or something more exotic). In the noncommutative geometric approach to constructing gauge theories the `spectral action' is most often used to describe dynamics. The spectral action  is introduced, in part, because it is difficult to construct dynamics by making explicit use of geometric objects like noncommutative differential forms. While intriguing and extremely beautiful, the spectral action  provides an effective description of nature, and is the setting within which the incorrect Higgs mass  prediction was made~\cite{carlm,Knecht_2006}. The calculus constructed in this paper presents a necessary step towards an explicit and more direct construction of dynamics.

The next immediate steps in fully constructing the internal Jordan geometries that  most closely capture the underlying structure of the  standard model of particle physics and viable extensions such as the Pati-Salam model are:
\begin{enumerate}
	\item \textbf{Discrete spaces:} It will  be necessary to consider finite Jordan coordinate algebras, 	which are semisimple, and in particular   of the form:
	\begin{align}
	A_{F} = \underbrace{H_n(\mathbb{C})\oplus...\oplus H_n(\mathbb{C})}_{`m'~copies},
	\end{align}
	with inner derivation algebra  $\mathrm{Der}(A_F) = su(n)\oplus...\oplus su(n)$. For coordinate algebras of this form we will need to generalize:
	\begin{itemize}
		\item \textbf{Discrete connections:} The algebra $A_F$ coordinatizes an `m' point (state) space with non-trivial finite internal structure at each point, which is a priori  completely disconnected. In order to `connect' such a  space, the notion of a discrete connection that relates the `m' (identical) factors  will be required. Higgs fields will arise in this way `connecting', for example, the chiral sectors of the models we are interested in.

		\item \textbf{Discrete calculus:} The derivation based calculus~\cite{Dubois_Violette_2016,Carotenuto_2018} is no longer appropriate when considering discrete geometries with disconnected sectors that are not smoothly related to one another by derivations. Connes has developed a `cochain' based generalization of differential forms appropriate for the discrete, noncommutative
		setting~\cite{Connes1994}. The corresponding `cyclic' cohomology only makes sense, however, for associative coordinate
		algebras, and so a generalization will need to be made that incorporates the benefits of both approaches. Generalizations have already been developed in very special cases, including for the octonion algebra~\cite{Akrami_2004,Beggs_2010}, and for 		Hom-associative algebras~\cite{Hassanzadeh_2015}, however there is as yet no general construction appropriate for Jordan algebras.

	\end{itemize}
	
		\item \textbf{Clifford representation of forms:} Much of the  predictive power of the noncommutative geometry that most closely resembles the standard model of particle physics derives from the geometric `rules' or axioms imposed on the finite input data. In practice, many of these rules have
been `ported across' from Riemannian spin geometry, where they derive naturally when considering the Clifford
action of forms on spinors. In the noncommutative setting, however, no clean analogue of a Clifford action of forms has so far been developed. As a result there is almost a complete
disconnect between the symmetries, signature, and scalar representations in these theories, which limits their
predictive power. A key step will be  to develop a `Clifford' representation of forms appropriate
for the Jordan geometries of interest. Once an appropriate notion of Clifford action is developed, it will be possible to explore whether the internal symmetries of a geometry correspond exclusively to the automorphisms of the input coordinate
algebra, or if there is some internal analogue of local Lorentz symmetry to be accounted for, as advocated
for example in~\cite{Chamseddine_2016} (and therefore what, if any, relationship exists between the symmetries and signature of an
internal space). Furthermore, another key exploration will be to determine whether the `multiplicity' required
in order to accommodate the internal analogue of a `Clifford' representation will lead to a prediction for the
number of fermion generations, or whether it will provide a stronger handle on the form of the fermion mass
matrices.

\end{enumerate}

\section*{Acknowledgements}
We  would like to thank  Lashi Bandara, John Barrett, Fabien Besnard, Latham Boyle, Branimir Cacic, Ludwik Dabrowski, Michel Dubois-Violette, Laurent Freidel,
 Hadi Godazgar, John Huerta, Axel Kleinschmidt, Giovanni Landi, Mathew Langford, Fedele Lizi, Matilde Marcolli, Pierre Marinetti, Fred Shultz, and Andrzej Sitarz
  for helpful conversations. This research was supported by an individual research grant from the Deutsche Forschungsgemeinschaft (project number 392887939), the European Research Council via ERC Consolidator Grant CoG 772295 ``Qosmology'', and the Max-Planck Institute.

\appendix

\section{The internal grading of the `noncommutative' standard model}
\label{gamma}

In equation~\eqref{mataction} we introduced the representation of the internal noncommutative coordinate algebra that most closely captures the fermion and gauge content of the standard model. We further introduced a finite charge conjugation operator $J_F$, which acts on elements of the Hilbert space by Hermitian conjugation. Here we introduce the finite grading operator  on elements $h\in H_F$ as $\gamma_F h = XhX$, where
\begin{align}
X= \left(\begin{tabular}{c c | c c}
\multirow{2}{*}{$+\mathbb{I}_2$} & \multirow{2}{*}{}& \multirow{2}{*}{}& \multirow{2}{*}{}\\
& & & \\
\multirow{2}{*}{}& \multirow{2}{*}{$-\mathbb{I}_2$}& \multirow{2}{*}{}& \multirow{2}{*}{}\\
& & & \\
\hline
\multirow{2}{*}{} & \multirow{2}{*}{}& \multirow{2}{*}{$\mathbb{I}_2$}& \multirow{2}{*}{}\\
& & & \\
\multirow{2}{*}{}& \multirow{2}{*}{}& \multirow{2}{*}{}& \multirow{2}{*}{$\mathbb{I}_2$}\\
& & & \\
\end{tabular}\right).
\end{align}

\section{Representations and unimodularity}
\label{appenA}
Let $A$ be an algebra defined over a field $\mathbb{F}$, and let $H$ be a vector space over $\mathbb{F}$. A bi-representation $\pi$ of $A$ on $H$  (or, equivalently, a bi-module
$H$ over $A$) is a pair of $\mathbb{F}$-bilinear products $f\Psi \in H$ and $\Psi f \in H$ ($f \in A, \Psi \in H$)\cite{Schafer1966,Boyle_2014,boyle2016new,Farnsworth_2015}. This definition of a bi-representation of $A$ on $H$ is equivalent to the definition of a new algebra
\begin{align} 
B = A \oplus H,
\end{align} 
with the product between elements of $B$ $(b = f + \Psi$ and $b'
= f'+ \Psi')$ given by
\begin{align}
bb' = ff' + f\Psi' + \Psi f', 
\end{align}
where $ff'\in A$ is the product inherited from $A$, while $f\Psi'\in H$ and $\Psi f' \in H$ are the products
inherited from $\pi$, and $\Psi\Psi' = 0$. We call such an algebra an "Eilenberg algebra".  

Notice that the above definition  has not assumed anything about the associativity of either $A$ or $B$. If however we assume that $B$ is associative, then we precisely recover the traditional  definition of an ordinary associative bi-representation of an associative algebra $A$ on $H$~\cite{Boyle_2014}. If on the other hand $A$ is a Jordan algebra, then we define its Jordan representation on
$H$ by taking $B$ to also be a Jordan algebra\cite{Boyle_2014,farnsworth2013nonassociative}. As an example consider the algebra $A = M_n(\mathbb{C})$ represented as matrices on $\mathbb{C}^{2n}$ by the following action:
\begin{align}
L_f\Psi &= \begin{pmatrix}f &\\
& 0
\end{pmatrix}\begin{pmatrix}\psi_L\\
\psi_R
\end{pmatrix},&
R_f\Psi &= \begin{pmatrix}0 &\\
& f^T
\end{pmatrix}\begin{pmatrix}\psi_L\\
\psi_R
\end{pmatrix},
\end{align}
for $f\in A$. The algebra $A$ is naturally involutive, with the involution given by Hermitian conjugation. We can similarly turn $B=A\oplus H$ into an  associative Eilenberg $*$-algebra by equipping $H$ with the following involution
\begin{align}
J = \begin{pmatrix}
& \mathbb{I}_n\\
\mathbb{I}_n
\end{pmatrix}\circ c.c,
\end{align}
where `$c.c$' denotes complex conjugation. 

The symmetries of a geometry correspond to the automorphisms of its coordinate algebra representation. In the above example the coordinate algebra representation is expressed in terms of an associative Eilenberg $*$-algebra $B = A+H$. The inner $*$-derivations on $B$ are then of the form $\delta_a = L_a - R_a$, where $a^*=-a\in A$ (the element $a$ can not  be taken from $H$, because automorphisms must be invertible). Remarkably, one finds that the inner $*$-derivation algebra of $B$ is larger than that of $A$, and is given by $u(n)$ (rather than $su(n)$). The generator $T=\delta_{i\mathbb{I}}$, while acting trivially on $A$, does not act trivially on $H$. This feature leads to  
an additional $U(1)$ gauge symmetry that is not observed experimentally, when constructing the  noncommutative geometry that  most closely captures the details of the standard model of particle physics. In order to remove the additional unwanted $U(1)$ symmetry, the so-called `unimodularity condition' is imposed, which is the statement that one should only take into account gauge generators that are traceless.

The `unimodularity condition' appears ad-hoc and poorly motivated. Remarkably, it appears as though this problem may be avoided in the Jordan setting. If in our above example,  one restricts to the maximal Jordan subalgebra $B_{sa} = A_{sa}+ H_{sa}$, equipped with the symmetrized product given in equation~\eqref{JordanprodZ}, then one obtains a Jordan representation of the real Jordan algebra $A_{sa} = H_n(\mathbb{C})$ on the vector space $\mathbb{R}^{2n}$. In this case the inner derivations on $B_{sa}$ are of the form $\delta_{ab} = [L_a,L_b]$, where $a,b\in A_{sa}$. It is an instructive (and not too difficult) exercise to show that the inner derivation elements on $B_{sa}$ form the Lie algebra $su(n)$. Every derivation of a semisimple algebra (that is, the direct sum of simple algebras) with a unit over a field of characteristic zero is inner, and as a result we know that all of the derivations of $A_{sa} = H_n(\mathbb{C})$ are inner (and form the Lie algebra $su(n)$). Notice however that the Eilenberg algebra $B_{sa} = A_{sa}\oplus H_{sa}$ is no longer semisimple, and as a consequence it  may have a larger set of symmetries. The full algebra of  $*$-derivations on $B_{sa}$ is in fact given by $u(n)$. In this case however, the additional $U(1)$ generator $T$ is explicitly distinguished from the inner $*$-derivations, and appears as an outer derivation.

\section{Differential calculi}
\label{AppendB}
Our aim in this section is to show that the differential graded algebra constructed over the Jordan algebra of $n\times n$ Hermitian matrices does indeed satisfy each of the properties outlined in section~\ref{diffstructure}. We begin by reminding the reader of the following well known  trace identities\cite{Kaplan,MacFarlane,Azcarraga,Fadin}:
\begin{alignat}{3}
Tr[F_i]&=0, & Tr[D_i]&=0, &  Tr[F_iF_j]&= n\hat{\delta}_{ij},\nonumber \\
Tr[D_iD_j]&= \frac{n^2-4}{N}\delta_{ij},&
Tr[F_iD_j]&= 0, &
Tr[F_iF_jF_k]&=i\frac{n}{2}f_{ijk},\nonumber\\
Tr[D_iF_jF_k]&= \frac{n}{2}d_{ijk}, &
Tr[D_iD_jF_k]&=i\frac{n^2-4}{2n}f_{ijk}, & 
Tr[D_iD_jD_k]&=  \frac{n^2-12}{2n}d_{ijk}, \nonumber\\
Tr[F_iD_jF_kD_l]&= \frac{1}{2}(\hat{\delta}_{ik}\hat{\delta}_{jl}-\hat{\delta}_{ij}&\hat{\delta}_{kl})+\frac{n}{4}(d_{iln}d_{jkn}&+f_{iln}f_{jkn}),\hspace{.21cm}&\hspace{.31cm}&\nonumber\\
Tr[F_iF_jD_kD_l] &=\frac{1}{2}(\hat{\delta}_{ij}\hat{\delta}_{kl}-\hat{\delta}_{ik}&\hat{\delta}_{jl})+\frac{n^2-8}{4n}f_{iln}&f_{jkn}+\frac{n}{4}d_{iln}d_{jkn},\hspace{.21cm}&\hspace{.31cm}&\nonumber\\
Tr[F_iD_jD_kD_l]&=i\frac{2}{n}f_{iln}d_{jkn}~+&i\frac{n^2-8}{4n}f_{ijn}d_{kln}+&\frac{i}{4}d_{ijn}f_{kln}    & &\nonumber\\
Tr[F_iF_jF_kD_l]&=i\frac{n}{4}(d_{iln}f_{jkn}~-&f_{iln}d_{jkn}), \hspace{1.3cm}  & & &\nonumber
\end{alignat}
together with the identities
\begin{subequations}
	\begin{align}
	0&=f_{ije}f_{ekl}+ f_{kje}f_{iel}+f_{lje}f_{ike},\label{jacobs}\\
0&=f_{ijn}d_{kln} +f_{iln}d_{jkn}+f_{jkn}d_{lin}\label{cycleYO}, \\
 f_{ije}f_{jke}&=\frac{2}{n}(\delta_{ik}\delta_{jl}-\delta_{il}\delta_{jk})+ d_{ikn}d_{jln} - d_{jkn}d_{iln}\label{symeq} 
\end{align}
\end{subequations}
where we have defined $(F_i)_{jk} =-if_{ijk}$ and $(D_i)_{jk} =d_{ijk}$. Equation \eqref{jacobs} is nothing other than the well known Jocobi identity. Our first goal will be to show that the basis of derivation elements $\delta_j\in \mathrm{Der}(A_{sa})$ and dual forms $E^i\in \Omega^1 A_{sa}$ defined in equations \eqref{Jordder} and \eqref{dualbasisJ} respectively,  satisfy the condition $E^i[\delta_j] = \hat{\delta}_{j}^i$. To see this, we make use of equation \eqref{derjoract}, together with Jordan product expressed as the symmetrization of the matrix product given in Eq.~\eqref{matprod} to  write:
\begin{align}
E^i[\delta_j] &= \frac{16}{n}f^{kli}(\lambda_l\circ\lambda^n)\circ(\lambda_k\circ (\delta_j\lambda_n))\nonumber\\
&=-\frac{4}{n^3}f^{kli}f_{jlk} +\frac{4}{n^2}f^{kli}(f_{ljn}d_{k}^{\phantom{k}nm}+f_{njk} d_{l}^{\phantom{l}nm})\lambda_m+\frac{4}{n}f^{kli}f_{njf} d_{l}^{\phantom{l}nh}d_{k}^{\phantom{k}fm} \lambda_h\lambda_m\nonumber\\
&=\frac{4}{n^3}Tr[F^iF_j] +\frac{4}{n^2}(Tr[F^iF_jD^k]+Tr[F^iD^kF_j])\lambda_k
+\frac{4}{n}Tr[F^iD^lF_jD^k] \lambda_l\lambda_k\nonumber\\
&=\frac{4}{n^2}\hat{\delta}_{j}^i +\frac{4}{n}d^{ik}_{\phantom{ki}j}\lambda_k
+\frac{2}{n}[\frac{n}{4}(d^{ikn}d_{jln}+f^{ikn}f_{ljn})-\frac{1}{2}(\hat{\delta}_{l}^i\hat{\delta}_{j}^k-\hat{\delta}_{j}^i\hat{\delta}_{l}^k)] (\frac{1}{n}\hat{\delta}_{k}^l + d_{k}^{\phantom{k}ln}\lambda_n)\nonumber\\
&=\frac{4}{n^2}\hat{\delta}_{j}^i +\frac{4}{n}d^{ik}_{\phantom{ik}j}\lambda_k
+\frac{2}{n^2}[\frac{n}{4}Tr[D^iD_j-F^iF_j]-\frac{1}{2}(\hat{\delta}_{j}^i-\hat{\delta}_{j}^i\hat{\delta}_{l}^l)  ]\nonumber\\
&\qquad+\frac{2}{n}[\frac{n}{4}Tr[D^{i}D_{j}D^{n}-F^{i}F_{j}D^{n}]-\frac{1}{2}d^{in}_{\phantom{in}j}]  \lambda_n\nonumber\\
&=
\hat{\delta}_{j}^i\label{dualresult},
\end{align}
as required. 

Our next goal will be to show that $dE^i = \frac{1}{2}f_{\phantom{i}jk}^i E^j\wedge E^k$. Before doing so however, we note that in deriving this result we will not  make use of the anti-symmetry of the wedge product itself. To remind the reader of this fact we explicitly replace the usual wedge product `$\wedge$' with an abstract product `$\bullet$'  throughout, reinstating the wedge product only at the end.  We then show  that the nil-potency condition $d^2=0$, implies the anti-symmetry of the product between forms, just as occurs in Riemannian geometry. We  begin by  making use of equation \eqref{dualbasisJ}, together with the symmetrization of the matrix  product given in Eq.~\eqref{matprod} to  write:
\begin{align}
dE^i &= \frac{8}{n}f^{jki}d[(\frac{1}{n}\hat{\delta}_{kl} + d_{kln}\lambda_n)\circ(\lambda_j \circ d\lambda_l)]\nonumber\\
  &= \frac{8}{n}f^{jki}\left[\underbrace{d_{klp}(\delta_m\lambda_p)\circ(\lambda_j\circ\delta_n\lambda_l)}_{(A)}+\underbrace{\frac{1}{n}\delta_m\lambda_j\circ \delta_n\lambda_k}_{(B)}+\underbrace{d_{lkp}\lambda_p\circ(\delta_m\lambda_j\circ \delta_n\lambda_l)}_{(C)}\right]E^m\bullet E^n\label{startproof}. 
\end{align}
We break equation~\eqref{startproof} down into three more manageable parts, and address each separately. We begin with the first term labelled `A':
\begin{align}
(A)&= \frac{8}{n}f^{jki}d_{klp}(\delta_m\lambda_p)\circ(\lambda_j\circ\delta_n\lambda_l)E^m\bullet E^n\nonumber\\
&=\frac{2}{n}f^{jki}d_{klp}f_{mps}\left[\frac{2}{n}f_{nlj}\lambda_s +f_{nlt}d_{jtq}(\frac{1}{n}\hat{\delta}_{sq}+d_{sqr}\lambda_r)\right]E^m\bullet E^n\nonumber\\
&=\frac{2}{n}\left[\frac{2}{n}f_{pms}Tr[F_iD_pF_n]\lambda_s+\frac{1}{n}f^{ijk}Tr[F_mD_bF_nD_k]-f_{mps}d_{sqr}Tr[F^iD_pF_nD_q]\lambda_r\right] E^m\bullet E^n\nonumber\\
&=\frac{2}{n}\left[f_{pms}d_{ipn}\lambda_s-\frac{1}{2n}f^{imn}+\frac{1}{4}f^{ijk}(f_{mce}f_{nej}+d_{mke}d_{nej})\right] E^m\bullet E^n\nonumber\\
&\qquad+\frac{2}{n}\left[\frac{1}{2}f_{mis}d_{snr}-\frac{n}{4}(f_{iqe}f_{nep}f_{mps}d_{rsq}+f_{mps}d_{rsq}d_{iqe}d_{nep})\right]\lambda_r E^m\bullet E^n\nonumber\\
&=\frac{2}{n}\left[f_{smr}d_{isn}\lambda_r-\frac{1}{2n}f^{imn}-\frac{i}{4}Tr[F^iF_mF_n-F^iD_mD_n]\right] E^m\bullet E^n\nonumber\\
&\qquad+\frac{2}{n}\left[\frac{1}{2}f_{mis}d_{snr}+\frac{n}{4}iTr[F_iF_nF_mD_r-F_mD_rD_iD_n]\right]\lambda_r E^m\bullet E^n\nonumber\\
&=0.
\end{align}
We next address the second term in equation~\eqref{startproof} labelled `B':
\begin{align}
(B)&= \frac{8}{n^2}f^{jki}\delta_m\lambda_j\circ \delta_n\lambda_kE^m\bullet E^n\nonumber\\
&=\frac{4}{n^2}f^{jki}f_{mjs}\left[\frac{1}{n}f_{nks}f_{nct}d_{stp}\lambda_{p}\right]E^m\bullet E^n\nonumber\\
&=\frac{4}{n^2}\left[\frac{i}{n}Tr[F^iF_nF_m]+iTr[F^iF_nD_pF_m]\lambda_p\right]E^m\bullet E^n\nonumber\\ 
&=\frac{1}{n}\left[\frac{2}{n}f_{imn}+(f_{mpe}d_{ine}-f_{ine}d_{mpe})\lambda_p\right]E^m\bullet E^n.
\end{align}
Finally, we address the third term in equation~\eqref{startproof} labelled `C':
\begin{align}
(C)&= \frac{8}{n}f^{jki}d_{lkp}\lambda_p\circ(\delta_m\lambda_j\circ \delta_n\lambda_l)E^m\bullet E^n\nonumber \\
&=\frac{4}{n}f^{jki}d_{klp}f_{mjs}\left[\frac{1}{n}f_{nls}\lambda_p+\frac{1}{2}f_{nlt}d_{stq}(\frac{1}{n}\delta_{pq}+d_{pqr}\lambda_r)\right]E^m\bullet E^n\nonumber\\
&=\frac{2}{n}\left[\frac{2}{n}iTr[F_nF_mF^iD_p]\lambda_p+\frac{1}{n}Tr[F_mD_tD_dF^i]f_{ndt}+if^{jki}f_{mbs}Tr[F_nD_sD_rD_k]\lambda_r\right]E^m\bullet E^n\nonumber\\
&=\frac{1}{n}\left[f_{npe}d_{mie}-f_{mie}d_{npe}+\frac{1}{n}f_{mni}+
\frac{n^2-8}{2n^2}f_{ile}f_{met}f_{ntl}-\frac{1}{2}d_{ile}d_{met}f_{ntl}\right]E^m\bullet E^n \nonumber\\
&+f^{ijk}f_{mjs}\left[\frac{4}{n^2}(f_{nse}d_{rek}+d_{rse}f_{nek})\lambda_r+\frac{1}{2}(d_{nse}f_{rek}-f_{nse}d_{rek})\lambda_r\right]E^m\bullet E^n \nonumber\\
&=\frac{1}{n}\left[f_{npe}d_{mie}-f_{mie}d_{npe}+\frac{1}{n}f_{mni}-i
\frac{n^2-8}{2n^2}Tr[F_iF_mF_n]-i\frac{1}{2}Tr[F_nD_iD_m]\right]E^m\bullet E^n \nonumber\\
&+\left[\frac{4}{n^2}iTr[F^iF_mF_nD_r + F^iF_mD_rF_n]\lambda_r-\frac{i}{2}Tr[F^iF_mF_nD_r-F^iF_mD_nF_r]\lambda_r\right]E^m\bullet E^n \nonumber\\
&=\frac{1}{n}\left[f_{nre}d_{mie}\lambda_r-f_{ime}d_{nre}\lambda_r+\frac{n}{2}f_{mni}-\frac{2}{n}f_{mni}\right]E^m\bullet E^n, 
\end{align}
where we have made significant use of equation \eqref{cycleYO}. The right hand side of equation~\eqref{startproof} is then given  by the summation of terms `A', `B' and `C':
\begin{align} 
dE^i = (A)+(B)+(C) = \frac{1}{2}f_{mni}E^m\bullet E^n\label{dONe}. 
\end{align}

We may replace the abstract product `$\bullet$' in equation \eqref{dONe} by the wedge product `$\wedge$', and indeed, the nilpotency condition $d^2=0$, implies that the product between two basis forms $E^i, E^j\in \Omega^1 A_{sa}$, must  be  anti-symmetric. To see this, we begin by assuming an abstract product between forms denoted by `$\bullet$'. We then have
\begin{align}
0 = d^2\lambda_i &= d(\delta_j\lambda_i E^j)\nonumber\\
&=f_{jik}f_{l}^{\phantom{l}kf}\lambda_f E^l\bullet E^j -f_{ji}^{\phantom{ji}k} \lambda_k dE^j\nonumber\\
&=\frac{1}{2}(f_{jik}f_{l}^{\phantom{l}kf}-f_{lik}f_{j}^{\phantom{j}kf})\lambda_f E^l\bullet E^j+\frac{1}{2}(f_{jik}f_{l}^{\phantom{l}kf}+f_{lik}f_{j}^{\phantom{j}kf})\lambda_f E^l\bullet E^j -f_{ji}^{\phantom{ji}k} \lambda_k dE^j\nonumber\\
&=f_{ji}^{\phantom{ij}k}\lambda_k( \frac{1}{2}f_{le}^{\phantom{le}j} E^l\bullet E^e -  dE^j)+\frac{1}{2}(f_{jik}f_{l}^{\phantom{l}kf}+f_{lik}f_{j}^{\phantom{j}kf})\lambda_f E^l\bullet E^j\nonumber\\
&=\frac{1}{2}(f_{jik}f_{l}^{\phantom{l}kf}+f_{lik}f_{j}^{\phantom{j}kf})\lambda_f E^l\bullet E^j,
\end{align}
where in the fourth line we have made use of the Jacobi given in equation \eqref{jacobs}, and in the last line we have made use of the result given in equation \eqref{dONe}. Finally, the last line is identically zero if we replace the  abstract product `$\bullet$' with one that is  anti-symmetric, such as the wedge product `$\wedge$'. The wedge product together with the graded Leibniz rule given in equation \eqref{DJord01} further ensure the that the  exterior derivative squares to zero on all higher order forms. In particular: 
\begin{align}
d^2E^i&= \frac{1}{2}f_{mni}d[E^m\wedge  E^n]\nonumber\\
&= \frac{1}{4}f_{mni}f_{stm}E^s\wedge E^t\wedge E^n - \frac{1}{4}f_{mna}f_{stn}E^m\wedge E^s\wedge  E^t\nonumber\\
&=-\frac{1}{2}[\frac{2}{n}(\delta_{is}\delta_{nt}-\delta_{it}\delta_{ns})+ d_{ism}d_{ntm} - d_{nsm}d_{itm}]E^s\wedge E^t\wedge E^n=0,\nonumber
\end{align}
where we have made use of equation \eqref{symeq}.


\bibliography{mybib}{}
\bibliographystyle{plain}

\end{document}